\documentclass[fleqn]{article}

\usepackage{graphics}

\usepackage{latexsym}

\begin{document}
\baselineskip 8truemm
\begin{titlepage}
\vspace{5mm}
\begin{center}
{\Large {\bf Remarks on the compactified six-dimensional model of
a particle} }
\vspace{6mm} \\
{\bf J. Syska}\footnote{jacek@server.phys.us.edu.pl }\\
\vspace{5mm}
{\sl Department of Field Theory and Particle Physics, Institute of Physics,} \\
{\sl University of Silesia, Uniwersytecka 4, 40-007 Katowice, Poland}\\
\noindent

\setcounter{equation}{0} \vspace{5 mm} ABSTRACT
\end{center}
\vspace{2 mm}

Non-homogeneous gauge ground state solutions in a six-dimensional
gauge model in the presence of non-zero extended fermionic charge
density fluctuations are reviewed and fully reinterpreted.

\vspace{20 mm}

Submitted to Physical Review D on 5 November 2001.


\vspace{2 mm} \vfill \vspace{5 mm}
\end{titlepage}

\section{Introduction}
\label{torus}

The goal of this paper is to present a simple, self-consistent
example of trivially coupled fields in a six-dimensional
spacetime with compactified internal dimensions (see
Section~\ref{torus-intr}). It is devoted mainly to the existence
of the nontrivial topological gauge ground field configuration
\cite{Dziekuje, Boson} on the internal space (torus
\cite{Manka-Syska}). The basic result of this paper (see
Sections~4 and~5) is the appearance of the gauge field as a
ground state solution (see Section~\ref{ground state}) of coupled
Dirac and Maxwell wave equations.
Because of vanishing current density fluctuations in the internal
space the screening current problem \cite{Dziekuje}, 
\cite{Aitchison-Hey} in the internal space is avoided.


\section{A short introduction to N-circles compactification}
\label{torus-intr}

Since the discovery of the cancellation of infinities in
superstring models, there has been a renewed interest in
higher-dimensional gauge theories. To begin, let us study a
$(4+N)$ dimensional Minkowskian spacetime with one temporal
dimension and $(3+N)$ spatial dimensions, equipped with the metric
tensor $\eta_{\mu \nu} = diag(+1,-1,...,-1)$. Now, consider a
Poincare invariant model in this spacetime - for example $\lambda
\phi^{4}$ scalar model given by action
\begin{equation}
\label{action-4-N}
{\cal S} = \int d^{4+N} x \left\{ \frac{1}{2}
\partial_{\alpha}\phi \partial^{\alpha}\phi - \frac{1}{2} \mu_{0}^{2} \phi^{2} -
\frac{1}{4!} \lambda_{0} \phi^{4} \right\} \; .
\end{equation}

To construct a classical model, we choose $N$ supplementary space
dimensions as circles of circumferences $L_{1},...,L_{N}$
respectively \cite{Cremmer-Scherk}. This assumption is equivalent
to periodic boundary conditions for the $\phi$ field:
\begin{equation}
\label{boundary} \phi(x_{3+i} + L_{i}) = \phi(x_{3+i}) \; .
\end{equation}

Now, me make a Fourier decomposition of the $\phi$ field:
\begin{equation}
\label{phi-decomposition} \phi(x_{\mu}, x_{3+i}) =
\frac{1}{(L_{1} \cdot ... \cdot L_{N})^{\frac{1}{2}}} \sum_{ \{
n_{i} \} } \phi_{ \{ n_{i} \} } \exp(2 \pi i \sum_{i=1}^{N}
\frac{x_{3+i} n_{i}}{L_{i}}) \; ,
\end{equation}
where the coefficients $\phi_{ \{ n_{i} \} }$ are fields
depending on the first $4$ spacetime dimensions $x^{\mu}$ only.
They fulfill the condition of being "real" $\phi_{ - \{ n_{i} \}
}^{*} = \phi_{ \{ n_{i} \} }$.

Integrating Eq.(\ref{action-4-N}) over the additional coordinates
$x_{3+i}$ from $0$ to $L_{i}$ one \cite{Scherk} easily obtains
the action
\begin{eqnarray}
\label{action-4} {\cal S} &=& \int d^{4} x \left\{ \frac{1}{2}
\partial_{\alpha}\phi_{ \{ n_{i} \} }^{*}
\partial^{\alpha}\phi_{ \{ n_{i} \} } -
\frac{1}{2} m_{ \{ n_{i} \} }^{2}
\phi_{ \{ n_{i} \} }^{*} \phi_{ \{ n_{i} \} } \right. \nonumber \\
&-& \left. \frac{\lambda}{4!} \phi_{ \{ n^{1}_{i} \} } \phi_{ \{
n^{2}_{i} \} } \phi_{ \{ n^{3}_{i} \} } \phi_{ \{ n^{4}_{i} \} }
\delta (n^{1}_{i} + n^{2}_{i} + n^{3}_{i} + n^{4}_{i}) \right\}
\; ,
\end{eqnarray}
where
\begin{eqnarray}
\label{lambda-4} \lambda = \frac{\lambda_{0}}{(L_{1} \cdot ...
\cdot L_{N})}
\end{eqnarray}
and
\begin{eqnarray}
\label{m-ni-4} m_{ \{ n_{i} \} }^{2} = \mu_{0}^{2} + 4 \pi^{2}
\sum_{i=1}^{N} \frac{n_{i}^{2}}{L_{i}^{2}} \; .
\end{eqnarray}
So, the model with compactified internal dimensions has infinite
number of states with an increasing ladder of masses. The formula
given by Eq.(\ref{m-ni-4}) was obtained, in the case of $N = 2$
and $\mu_{0} = 0$, for particles and solitons in supersymmetric
gauge theory.

The dimensional reduction is based on $L_{i} \rightarrow 0$
transition (for fixed $\lambda$). Then only one field, $ \phi_{
\{ n_{i} = 0 \} } \equiv \phi$, has the finite mass. Therefore, we
receive the usual $\lambda \phi^{4}$ model in $4$-dimensional
spacetime:
\begin{eqnarray}
\label{action-R} {\cal S} \rightarrow {\cal S}_{R} = \int d^{4} x
\left\{ \frac{1}{2} \partial_{\alpha}\phi \partial^{\alpha}\phi -
\frac{1}{2} \mu_{0}^{2} \phi^{2} - \frac{\lambda}{4!} \phi^{4}
\right\} \; .
\end{eqnarray}

In the above proposal the dimensional reduction erases any trace
of the hidden model of the world.

A little bit complicated example is obtained when we start with
the Maxwell field in $(4+N)$ dimensions:
\begin{eqnarray}
\label{action-Maxw} {\cal S} = - \frac{1}{4} \int d^{4+N} x
F_{\alpha \beta} F^{\alpha \beta} \; ,
\end{eqnarray}
where
\begin{eqnarray}
\label{Maxw-field} F_{\alpha \beta} = \partial_{\alpha}A_{\beta}
- \partial_{\beta}A_{\alpha} \; .
\end{eqnarray}
Leaving aside the subsequent steps in compactification (analogous
to Eqs.(\ref{boundary})-(\ref{m-ni-4})), we suppose that
$A_{\alpha}$ ($\alpha = 0,1,...,N$) field does not depend on
$x^{3+i}$.
Now, the $A_{\alpha}$ field is decomposed an $A_{\mu}$ field
($\mu = 0,...,3$) and $\phi_{i}$ scalar fields ($i = 1,...,N$), so
$A_{\alpha} = (A_{\mu}, \phi_{1},...,\phi_{N})$ (see Section~{\bf
\ref{torus-model}}).

As a result of this procedure, the Poincare invariance of the
original $(4+N)$ dimensional theory remains hidden, and only the
Poincare group in $4$~spacetime dimensions and $O(N)$ group are
effective, with $O(N)$ singlet-$A_{\mu}$ and $O(N)$
vector-$\phi_{i}$.


\section{Introduction to the self-consistent theory of classical fields}
\label{ground state}

In the present paper the language of self-consistent theory of
classical fields is used. Hence in order to understand the model
in this broader context let us for a moment simplify our
considerations taking into account a real scalar field theory
model defined by the following Lagrangian density:
\begin{eqnarray}
\label{scalar-Lagr} {\cal L} = \frac{1}{2} \dot{\phi}^{2}(x,t) -
\frac{1}{2}(\nabla \phi(x,t))^{2} - V(\phi)
\end{eqnarray}
where $\dot{\phi} = \partial \phi/\partial t$, $\nabla \phi =
\sum_{i} \bar{i} \; \partial\phi/\partial x^{i}$ ($\bar{i}$ is
the versor).

$V(\phi)$ is a function of $\phi$ and the dependence on the
coupling constant $g$ is given by
\begin{eqnarray}
\label{V-phi} V(\phi) = \frac{1}{g^{2}} \tilde{V}(\chi), \; \; \;
\chi = g \phi
\end{eqnarray}
where $\tilde{V}$ is an even function independent of $g$.
Depending on the choice of the functional form of $\tilde{V}$ one
can consider various models. One of its realization is illustrated
on Figure~1.

The Hamiltonian density derived from the Lagrangian of
Eq.(\ref{scalar-Lagr}) is given by
\begin{eqnarray}
\label{scalar-Hamilt} {\cal H} = \frac{1}{2} \dot{\phi}^{2} +
\frac{1}{2}(\nabla \phi)^{2} + V(\phi)
\end{eqnarray}
Let $\phi_{0}$ be the scalar timeless field solution of the
equation of motion for the field $\phi$ in the ground state of
the system given by this Hamiltonian. We use the name of a {\it
scalar ground field} for the solution $\phi_{0}$.

If we are interested in the Lagrangian density
\begin{equation}
\label{lagr-el} {\cal L} = \bar{\Psi} ( \gamma^{\mu} i
\partial_{\mu} - m ) \Psi + J^{\mu} \, A_{\mu} - \frac{1}{4} \,
F_{\mu \nu} \, F^{\mu \nu} \, ,
\end{equation}
where $J^{\mu} = - e \bar{\Psi} \gamma^{\mu} \Psi $ is the
electron current density fluctuation and, $A_{\mu} $ is the total
electromagnetic field, four-potential $A_{\mu} = A^{e}_{\mu} +
A^{s}_{\mu} $, with the superscripts $e$ and $s$ standing for
external field and self field (which is adjusted by the radiative
reaction to suit the electron current and its fluctuations, see
\cite{Barut}), respectively, then, in the minimum of the
corresponding total Hamiltonian, the solution of the equation of
motion for $A^{s}_{\mu} $ is called {\it electromagnetic ground
field}. At this point a serious warning has to be given: {\it In
the present paper, as in Barut model \cite{Barut} also, the
Dirac-Maxwell coupled equations contain the wavefunction
$\Psi(x)$ which does not have the interpretation connected with
full charge density distribution, as in the original linear
Schr\"{o}dinger or Dirac equations, but it is connected with the
fluctuations of charge density distribution.}

More generally we use the name of a {\it boson ground field} for
a solution of an equation of motion for a {\it boson field} in
the ground state of a whole system of fields (fermion, gauge
boson, scalar) under consideration. This boson field is the self
field  (or can be treated like this) when it is coupled to a
source-"basic" field. 
By "basic" field we mean a wave (field) function which is proper
for a fermion, a scalar or a heavy boson. This concept of a wave
function and the Schr\"{o}dinger wave equation is dominant in
nonrelativistic physics of atoms, molecules and condensed matter.
In the relativistic quantum theory this notion had been largely
abandoned in favour of the second quantized perturbative Feynman
graph approach, although the Dirac wave equation is used
approximatively in some problems. What was done by Barut and
others was an extension of the Schr\"{o}dinger's "charge density
interpretation" of a wave function\footnote{ Electron is a
classical distribution of charge.} to a "fully-fledged"
relativistic theory. They implemented successfully this "natural
(fields theory) interpretation" of a wave function in many
specific problems with coupled Dirac and Maxwell equations (for
characteristic boundary conditions). But the "natural
interpretation" of the wave function could be extended to the
Klein-Gordon equation \cite{Dziekuje} coupled to Einstein field
equations, thus being a rival for quantum gravity in its second
quantized form. In both cases the second quantization approach is
connected with the probabilistic interpretation of quantum
theory, whereas the "natural interpretation" together with the
self field concept goes in tune with the deterministic
interpretation composing a relativistic, self-consistent field
theory.

To summarize: Depending on the model, the role of a self field
can be played by electromagnetic field \cite{bib B-K-1,bib
B-H,bib B-D,bib B,spontaneous,Unal,bib J,bib M}, boson $W^{+}
\cup W^{-}$ ground-field \cite{Dziekuje, Boson}, or by the
gravitational field (metric tensor) $g_{\mu \nu}$
\cite{Dziekuje}. The main law for arising of these self fields
would be taking the lead existence of "basic" fields.

\section{The model}
\label{torus-model}

Let us consider a six-dimensional field theory of a
"fermion"\footnote{Quotation mark denotes the fact that half spin
of the particle is observed from the outside only; generally
speaking inside a particle the Lorentz symmetry is broken}
particle in a curved spacetime.
\begin{equation}
\label{row_com1} {\cal L} = - \frac{1}{4} F^{a}_{M N} F^{a \, M
N} + i\bar{\psi} \Gamma^{M} \nabla_{M} \psi \, ,
\end{equation}
where $\psi$ is the "fermion" wave function\footnote{Connected
with the charge (matter) fluctuation density (see
Section~\ref{ground state})} of an extended particle coupled to
its self field with a field strength tensor $F^{a}_{M N}$ given by
\begin{equation}
\label{row_com8} F^{a}_{M N} = \partial_{M} W^{a}_{N} -
\partial_{N} W^{a}_{M} - g f_{abc} W^{b}_{M} \, W^{c}_{N}  \; .
\end{equation}
The covariant derivative $\nabla_{M}$ is
\begin{equation}
\label{row_com9} \nabla_{M} = \partial_{M} + \frac{1}{2}
\omega_{MAB} \Sigma^{AB} + i g W_{M} \; ,
\end{equation}
where
\begin{equation}
\label{row_com3} W_{M } = W^{a}_{M} T_{a} \; , \;\;\; \left[
T_{a}, T_{b} \right] = i f_{abc} \, T_{c}
\end{equation}
and
\begin{equation}
\label{row_com10} \Sigma^{AB} = \frac{1}{4} \left[ \Gamma^{A}, \,
\Gamma^{B} \right] \; .
\end{equation}
Indices $M,N,...,$ take the values 0,1,2,3 and 5,6.
$\omega_{MAB}$ is the spin connection. $\Gamma_{M} \equiv
e^{M}_{A} \Gamma^{A}$, where $\Gamma^{A}$ are $\gamma$ matrices
in the flat six-dimensional Minkowski spacetime. Generator
$T_{a}$ is an element of a Lie algebra {\bf q}.

We \cite{Manka-Syska} impose the geometry of the six-dimensional
spacetime to be the topological product of the flat
four-dimensional Minkowski (external) spacetime $(\eta_{\mu
\nu})$ (with $ \mu , \nu = 0,1,2,3 $), and the internal space
(with metric $g_{eh}, \; e,h = 5,6 $) which forms a generalized
torus. Therefore, the metric tensor can be factorized as
\begin{eqnarray}
\label{row_com11} g_{MN} = \pmatrix{ \eta_{\mu \nu } & 0 \cr 0 &
- g_{eh} }
\end{eqnarray}
with the two-dimensional internal part\footnote{The
observationally meaningful case with flat two-dimensional
internal manifold is discussed in \cite{Dar}.}
\begin{eqnarray}
\label{row_com12} g_{eh} = \pmatrix{  (R + \epsilon r
cos\vartheta)^{2}  & 0 \cr
 0 &  r^{2} } \; .
\end{eqnarray}

The ordinary torus can  be visualized as a two-dimensional surface
embedded in the three-dimensional Euclidean space with
coordinates given by
\begin{eqnarray}
\label{row_com13} \left\{\begin{array}{lll}
w^{1} =  ( R + r \, cos \vartheta) \, cos \phi \;\;  \\
w^{2} =  (R + r \, cos \vartheta) \, sin \phi  \;\; \\
w^{3} =  r \, sin \vartheta \; \; , \; \; \; \; \; \phi \in [0, 2
\: \pi ) \, ,
 \;\; \vartheta \in [0, 2 \pi ] \;
\end{array}
\right.
\end{eqnarray}
with $x^{5} = \phi$, $x^{6} = \vartheta$, which corresponds to
the case with $\epsilon = 1$. The case $R = 0$ and  $\psi \in [0,
2 \pi)$ and $\vartheta \in [0, \pi]$ corresponds to a sphere. In
general, even for the torus case ($\epsilon = 1$), the internal
manifold is not flat (see Appendix). Only for $\epsilon = 0$
($S^1 \times S^1 $ torus) the flat internal space is obtained.

The Lagrange density function of Eq.(\ref{row_com1}) leads to the
Dirac equation
\begin{eqnarray}
\label{row_com14} i \Gamma^{M} \nabla_{M} \psi = 0
\end{eqnarray}
with $\nabla_{M} $ given by Eq.(\ref{row_com9}) and $\Gamma_{M} =
e^{M}_{A} \Gamma^{A}  $ and to the Maxwell equation coupled with
the Dirac equation
\begin{eqnarray}
\label{row_com16} \frac{1}{\sqrt{- g}} D_{M} \left( \sqrt{- g}
F^{a MN} \right) = j^{a N}
\end{eqnarray}
with
\begin{eqnarray}
\label{row_com17} (D_{M} \phi)^a = \partial_{M} \phi^{a} - g
f_{abc} A^{b}_{M} \phi^{c}
\end{eqnarray}
where repers $e^{M}_{A}$ are explicitly given in Section~{\bf
\ref{compact}} and $j^{a N}$ is the fermionic current density
fluctuation
\begin{eqnarray}
\label{row_comj} j^{a N} = - g \bar{\psi} \Gamma^{N} T_{a} \psi
\; .
\end{eqnarray}

The fundamental result of this Section is that the gauge field
appears as a ground state solution of the coupled wave equations
(\ref{row_com14})-(\ref{row_com16}). Hence, according to our
notation, it is a ground field.

Let us now divide the gauge field $A_{M}$ into the ground state
component $a_{M} (y) \, \, (y = \{ x^{5},x^{6} \}$, $M \equiv m =
5,6)$ in the internal space (so we will not be interested in the
excited solutions in the internal space), and $\tilde{A}_{M} (x)
\, (x = \{ x^{0},x^{1},x^{2},x^{3} \}, M \equiv \mu = 0,1,2,3),$
which is the component in the four-dimensional spacetime:
\begin{eqnarray}
\label{row_com18} A_{M} (x,y) = \left\{\begin{array}{lll}
a_{m} (y) \neq 0 \, , \; \;  M \equiv m = 5,6 \; , \\
\tilde{A}_{\mu} (x) \, , \; \; M \equiv \mu = 0,1,2,3  \;\; .
\end{array}
\right.
\end{eqnarray}
\nopagebreak[4] This paper is devoted to the gauge ground field in
the internal space. \cite{Dziekuje} and \cite{Boson} are devoted
to the existence of the gauge ground fields (ground state
solutions) in the four dimensional spacetime.

Let us suppose that a ground field in the internal space exists
in some Lie algebra direction of the Cartan subalgebra $h \subset
q$ only. Then
\begin{eqnarray}
\label{row_com19}
a_{m} = a_{m}^{i} H_{i} , \, \, H_{i} \in h \nonumber \\
a_{m}^{i} = a_{m} n^{i} \, ,
\end{eqnarray}
where $n^{i}$ indicates a certain direction in the Cartan
subalgebra $h$.

Now, the ground state gauge field $a^{i}_{m}$  is a solution of
the Maxwell equations due to the Abelian nature of the Cartan
subalgebra. We have
\begin{eqnarray}
\label{row_com20} \frac{1}{\sqrt{- g}} \partial_{m} \left(
\sqrt{- g} \, f^{mn}_{C} \right) = j^{n}
\end{eqnarray}
with
\begin{eqnarray}
\label{row_com21} f^{mn}_{C}  =  g^{mn} g^{nr} f^{C}_{pr} \; , \;
\; \; f_{pr} = \partial_{p} a_{r} - \partial_{r} a_{p} \, .
\end{eqnarray}
Let us discuss the simplest model with current density
fluctuations $j^{n}$ in the internal space equal to
zero.\footnote{ This assumption will be represented by the
example of Section~\ref{chiral}.} Thus, we will avoid the
screening current problem (see \cite{Dziekuje, Boson}) in the
internal space.

There are two types of solutions, i.e. homogeneous and
non-homogeneous, with respect to the internal coordinates
$y^{m}$. The homogeneous solutions
\begin{eqnarray}
\label{row_com22} a_{\phi}^{C} = n  \; , \; \; \; \;
a_{\vartheta}^{C} = 0
\end{eqnarray}
could not be gauged away due to the non-trivial topological
structure of a torus.

There is also a non-homogeneous solution of the Maxwell equation
Eq.(\ref{row_com20})
\begin{eqnarray}
\label{row_com23} a_{\phi}^{C} (\vartheta) =  a_{0} \left[
\vartheta + \epsilon (r/R) sin
\vartheta \right] \; , \\
a_{\vartheta}^{C} (\vartheta) = 0 \, . \nonumber
\end{eqnarray}
This is a generalization of the monopole solution. Indeed, if we
change the torus to the sphere $S^{2}$, the monopole solution
will be obtained. As we shall see, this ground field in the
internal space is needed to obtain the chiral zero mode in the
four-dimensional spacetime.

\section{Six-dimensional Dirac equation}
\label{chiral}

Let us consider \cite{Manka-Syska} the Dirac equation
\begin{eqnarray}
\label{row_com24} i e^{M}_{A} \Gamma^{A} \nabla_{M} \psi = 0
\end{eqnarray}
with $\nabla_{M} $ given by Eq.( \ref{row_com9}), where
$e_{A}^{M}$ are defined so that
\begin{eqnarray}
\label{row_com25} g_{mn} = e_{m}^{a} e_{n}^{b} \delta_{ab} \, ,
\; \; \; e_{\mu}^{\alpha} = \delta_{\mu}^{\alpha} \, , \; \; \;
e^{a}_{m} e^{m}_{b} = \delta^{a}_{b} \, .
\end{eqnarray}
$\Gamma^{A}$ are $\gamma$ matrices in flat six-dimensional
Minkowski spacetime, defined as follows:
\begin{eqnarray}
\label{row_com26}
\Gamma^{\mu} = \gamma^{\mu} \otimes I \; \; ,  \; \\
\Gamma^{5} = - i \gamma^{5} \otimes \sigma^{2} \;\;\; ,
\Gamma^{6} = i \gamma^{5} \otimes \sigma^{1} \;\;\; , \Gamma^{7}
= \gamma^{5} \otimes \sigma_{3}  \, .
\end{eqnarray}
The Lorentz SO(2) generator $\Sigma^{56}$ is
\begin{eqnarray}
\label{row_com27} i \Sigma_{3} = \Sigma^{56} = \frac{1}{4} \left[
\Gamma^{5} , \Gamma^{6} \right] = \frac{1}{2} i I \otimes
\sigma_{3} \, .
\end{eqnarray}
The Riemannian spin connection on compact internal manifold is
\cite{Wess}
\begin{eqnarray}
\label{row_com28} \omega_{nab} (e) = \frac{1}{2} e^{m}_{a}
e^{k}_{b} \left( \Omega_{mnk} + \Omega_{nkm} - \Omega_{kmn}
\right)
\end{eqnarray}
with\footnote{\label{f-cyt-4} Connection $\Omega_{mnr}$ is a
gravitational analog of field strength tensor $F_{MN}$ in
electromagnetism.}
\begin{eqnarray}
\label{row_com29} \Omega_{mnr} = e^{a}_{r} \left( \partial_{m}
e_{an} - \partial_{n} e_{am} \right) \, .
\end{eqnarray}
The metric Eqs.(\ref{row_com11})-(\ref{row_com12}) give
\begin{eqnarray}
\label{row_com30} \omega_{112} (e) = - \omega_{121} (e) =
\epsilon \, sin \vartheta \, .
\end{eqnarray}
A more general case is found when we introduce the torsion. This
changes the connection $\omega_{mab} (e)$ to
\begin{eqnarray}
\label{row_com31} \omega_{mab} = - \omega_{mab} (e) + K_{mab} \; ,
\end{eqnarray}
where $K_{mab}$ is the contorsion. This will influence only
fermion field fluctuation. The introduction of contorsion
$K_{mab}$ changes the Christoffel connection $\Gamma^{k}_{mn}
(e)$ to an asymmetric quantity
\begin{eqnarray}
\label{row_com32} \Gamma^{k}_{mn} =  \Gamma^{k}_{mn} (e) +
K^{k}_{mn}
\end{eqnarray}
with $K^{k}_{mn} = g^{kl} e^{a}_{m} e^{b}_{n} K_{lab}$. This
also gives us the torsion
\begin{eqnarray}
\label{row_com33} S_{mnk} =  - K_{mnk} + K_{nmk} \, .
\end{eqnarray}
As we shall see, a non-trivial torsion will be necessary to
obtain the chiral zero mode of the fluctuation\footnote{ Hence
the mass of this fermion particle on the ground state does not
change. }. If we express the $2^3$ spinor components as the
product
\begin{eqnarray}
\label{row_com34} \psi (x,y) = \sum_{\lambda} \psi_{\lambda} (x)
\otimes \epsilon_{\lambda} (y)
\end{eqnarray}
of the bispinor $\psi_{\lambda} (x)$ component and two
$\epsilon_{\lambda} (y)$ spinor components, then the Dirac
equation can be expressed as\footnote{\label{s-f-absence} From
the four-dimensional perspective we are investigating a free
Dirac field on the ground state, thus the self field absence at
the right hand side of Eq.(\ref{row_com35}) is justified and the
bispinor $\psi_{\lambda} (x)$ is connected with the
"Schr\"{o}dinger" charge density distribution only. }
\begin{eqnarray}
\label{row_com35} i \gamma^{\mu} \partial_{\mu} \psi_{\lambda} +
\lambda \gamma^{5} \psi_{\lambda} = 0 \; ,
\end{eqnarray}
\begin{eqnarray}
\label{row_com36} \pmatrix{0 & \nabla^{+} \cr \nabla_{-} & 0 }
\epsilon_{\lambda} = \lambda \epsilon_{\lambda}
\end{eqnarray}
with
\begin{eqnarray}
\label{row_com37} \nabla_{+} = - \frac{1}{r} \partial_{\vartheta}
+ \frac{\left( - i \partial_{\phi} - g a_{0} [ \vartheta +
\epsilon (r/R) sin \vartheta ] - ( K_{112} + \epsilon sin
\vartheta)/2 \right)} {(R + \epsilon r cos \vartheta )} \; ,
\end{eqnarray}
\begin{eqnarray}
\label{row_com38} \nabla_{-} = - \frac{1}{r} \partial_{\vartheta}
- \frac{\left( - i \partial_{\phi} - g a_{0} [ \vartheta +
\epsilon (r/R) sin \vartheta ] + ( K_{112} + \epsilon sin
\vartheta)/2 \right)} {(R + \epsilon r cos \vartheta )} \; .
\end{eqnarray}
The chiral zero mode ($\lambda = 0$) exists only if we link the
gauge connection $a_{m}$ with the spin connection $\omega_{112}$
(the Riemannian spin connection $\omega_{112} (e)$ and contorsion
$K_{112}$)\footnote{So, we see the importance of the existence of
the ground field $a_{m}$. }. This will occur if
\begin{eqnarray}
\label{row_com39} - g \frac{r}{R} \, a_{0} = \frac{1}{2}
\end{eqnarray}
and
\begin{eqnarray}
\label{row_com40} - g a_{0} \, \vartheta = - \frac{1}{2} K_{112}
\, .
\end{eqnarray}

Substituting
\begin{eqnarray}
\label{row_com41} \epsilon_{\lambda} =  e^{i m \phi}
\varepsilon_{\lambda} (\vartheta)
\end{eqnarray}
we obtain
\begin{eqnarray}
\label{row_com42} \nabla_{+} = - \frac{1}{r} \partial_{\vartheta}
+ \frac{m - 2 g a_{0} \vartheta}{(R + \epsilon r cos \vartheta )}
\; ,
\end{eqnarray}
\begin{eqnarray}
\label{row_com43} \nabla_{-} = - \frac{1}{r} \partial_{\vartheta}
- \frac{m + \epsilon sin \vartheta}{(R + \epsilon r cos \vartheta
)} \; .
\end{eqnarray}

The zero mode $\varepsilon_{0}(\vartheta)$ condition
\begin{eqnarray}
\label{row_com44} Q \varepsilon_{0} = \bar{Q} \varepsilon_{0} = 0
\, ,
\end{eqnarray}
where
\begin{eqnarray}
\label{row_com45} Q = \pmatrix{0 & \nabla^{+} \cr 0 & 0 }  \; ,
\end{eqnarray}
\begin{eqnarray}
\label{row_com46} \bar{Q} = \pmatrix{0 & 0 \cr \nabla_{-} & 0 }
\end{eqnarray}
gives
\begin{eqnarray}
\label{row_com47} \varepsilon_{0} (\vartheta) = \left(
\matrix{\chi \cr 0 \cr} \right) \exp ( - \int d\vartheta \frac{m
+ \epsilon sin \vartheta } {(R + \epsilon r cos \vartheta )} ) \;
.
\end{eqnarray}

For $m = 0$, equation Eq.(\ref{row_com47}) gives
\begin{eqnarray}
\label{row_com48} \varepsilon_{0} (\vartheta) =
\left(\matrix{\chi \cr 0 \cr} \right) \left( R + \epsilon r cos
\vartheta \right) \; .
\end{eqnarray}

When we define the internal current density fluctuations as
\begin{eqnarray}
\label{row_com50} j^{i} \equiv \overline{\varepsilon_{0}}
\sigma^{i} \varepsilon_{0} \, , \; \; \overline{\varepsilon_{0}}
= \varepsilon_{0}^{T}  \, , \; \; i = 5,6 \, ,
\end{eqnarray}
then we see that
\begin{eqnarray}
\label{row_com51} j^{5} = j^{6} = 0
\end{eqnarray}
which is an important result, and vanishing of the current density
fluctuations $j^{i}$ in the Maxwell equation (see
Eq.(\ref{row_com20})) is justified.

Finally, as the six-dimensional chirality is defined by $i
\gamma^{5} \otimes \sigma^{3}$, it is easy to show that the zero
mode of the internal fermionic field fluctuation $\epsilon_{0}$
is chiral. This means that
\begin{eqnarray}
\label{row_com49} \sigma_{3} \epsilon_{0} = + \epsilon_{0}
\end{eqnarray}
has positive internal chirality.

When $R \rightarrow 0 \, (\vartheta \in [0,\pi], \phi \in [0, 2
\pi))$, the zero mode $\epsilon_{0}$ (see Eq.(\ref{row_com47}))
is exactly the same as for the sphere, and the contorsion
$K_{112}$ vanishes.

{\it Note}: It is crucial to distinguish between field wave
mechanics and field fluctuation wave mechanics. Because of this
distinction our model has given an answer to the existence of
$\epsilon_{0}$ internal fermionic field fluctuation and its
chirality, but is not able to describe the chiral modes for the
global ("Schr\"{o}dinger") internal fermionic field.

\section{Conclusions and Perspectives}

The goal of this paper was to present a model of a hypothetical
particle existing in the six-dimensional spacetime with two
compactified internal dimensions and with two coupled fields: the
fermionic field fluctuation and its self field. Both of them
compose the ground state of the system. Hence the coupled
Dirac-Maxwell wave equations had to be solved. As we saw, the
ground field $a_{m}$ (and the non-trivial torsion $S_{mnk}$) on
the internal space were necessary to obtain the chiral zero mode
of the fluctuation in four-dimensional spacetime as well as on
the internal torus.

The question remains: If one fermionic field fluctuation has
global chirality (or other quantum numbers) of its own, are the
quantum mechanical rules relevant for the description of the way
in which the global chirality of these fluctuations constitute the
global chirality (or its possible change) of an entire fermionic
particle? If partons are fluctuations inside the proton then this
question seems to be of extreme importance.

Although the model presented above touches in some respect the
problem of the structure of one extended particle,
it does not mean that the model is able to describe its
complicated subtleties. Nevertheless the language is worth to be
used and developed. It is the language of the self-consistent
field theory. Its four-dimensional elocution leads in the case of
the specific electroweek model \cite{Dziekuje, Boson, Marek-Jacek}
to the particle and astroparticle (gamma ray bursts) applications.

I hope that from the time of Barut and others \cite{bib B-K-1,bib
B-H,bib B-D,bib B,spontaneous,Unal,bib J,bib M}, who have
developed the idea of the self-consistent treatment of
source-field effects via the process of the radiation reaction, a
new model of a matter particle have begun to appear. This model
should be seen as a consequence of the previous (although not
self-consistent) Schr\"{o}dinger's ideas which concern the
interpretation of the wave function. This interpretation underwent
changes while he [Schr\"{o}dinger] developed his wave mechanics,
and "Instead of publishing just one final result, he revealed the
whole process of his search, the picture of his long wandering in
the darkness" \cite{Vlasov}. What lightened the darkness was an
electrodynamic interpretation of the wave function of the
electron. But, because Barut coupled the Dirac equation to the
Maxwell one hence the "Barut" wave function does not have an
interpretation connected with the full charge density
distribution, as in the original linear Schr\"{o}dinger equation,
but connected with the charge density distribution fluctuations.
This devious situation persists till now and we have the
"dualism" having not been able to accumulate the description of
Schr\"{o}dinger's states and Barut's fluctuations of these states
in one single picture.  Also quantum mechanics together with QED
(and QCD, and GSW model)\footnote{It is said that in quantum
field theory a matter particle is treated as a set of aggregated
quanta \cite{Teller}. The strict point-likeness of mentioned
quanta is supposed to be the corner-stone of the theory.} do not
offer such a consistent picture\footnote{In the beginning,
Schr\"{o}dinger could not see the connection with Heisenberg's
method, but within a short time he established \cite{Schrodinger}
the equivalence between the two (see also \cite{Dirac})
approaches. Some suggested that there is also a connection
between the self field approach and quantum field theory. Because
the self field picture is in line with the classical point of
view, where there are no infinite energy density zero-point
fluctuations, and the vacuum field is identically equal to zero,
this statement sounds strange. But the 1951 paper of Callen and
Welton on the fluctuation dissipation theorem showed that there
is an intimate connection between vacuum fluctuations and the
process of radiation reaction. The existence of one implies the
existence of the other. Finally, to quote Milonni \cite{bib M}:
"It seems ... that the generalization of these ideas ... may lead
us to view the vacuum field more as a formal artifice or
subterfuge than a "real" physical thing." (see also
\cite{Jaynes})}. But have the model been formulated we might have
said that there is the model of an elementary matter particle.

Once somebody postulated the view of a particle as a definite
material entity (for example, a charge distribution described by
a wave function) which has a self field of its own coupled to it.
Just here the self-consistent theory with the "natural
interpretation" of a wave function finds its important
application: {\it a "basic" field (the cornerstone of a particle)
extends everywhere in the space accessible for a wave function
determined by its equation of motion and at the same time a self
field coupled to this "basic" field propagates according to its
field equations (Maxwell's--for the electromagnetic self field,
Einstein (or a better one)--for the gravitational self
field\footnote{In my opinion the gravitation has to be obligatory
incorporated.}) with the velocity of light}. Now, this is the
final goal of the one particle physics which has to be achieved.

Hence from the mathematical point of view, a stable elementary
particle is a self-consistent time and space dependent solution
of field equations involved in its description, which moves in the
spacetime in the nonspreading way. Then, what does remain? The
solution of the initial-value problem for coupled equation of
motion and field equations. Some "simple" stationary,
self-consistent, non-perturbative case of the coupled
Klein-Gordon-Einstein equations is discussed in \cite{Dar}.
However the inclusion of the Einstein equation to the system of
equations procures big analytical complications, so in view of
the outward similarity of the Brans-Dicke model to the discussed
one, even the perturbative calculations of the kind of the
Brans-Dicke scalar-tensor theory could be useful.

Finally, we could do all self-consistent calculations in the
six-dimensional space. At this point the model presented before
as homogeneous in its space-time structure seems to be an
oversimplification of the matter particle nature. We can imagine
that our six-dimensional world could be compactified in a
non-homogeneous manner \cite{Dar}. In this picture the real
massless scalar "basic" field $\varphi$ (dilaton) forms a kind of
ground field (but not the self field). At this point the model
which is discussed in \cite{Dar, Dziekuje} raises a new issue of
the analysis. The solutions presented in \cite{Dar} are
parameterized by the parameter $A$ which has similar dynamical
consequences as the mass\footnote{$G$ is the gravitational
constant.} $M = A c^2/(2 G)$ --- its existence would be perceived
by an observer in the same way as invisible mass which could be
the extended "center" of a particle.

\section{Appendix: Metric tensor in the model}
\label{compact}

The metric  Eq.(\ref{row_com12})
\begin{eqnarray}
\label{row_comA1} g_{eh} = \pmatrix{  (R + \epsilon r
cos\vartheta)^{2}  & 0 \cr
 0 &  r^{2} }
\end{eqnarray}
gives
\begin{eqnarray}
\label{row_comA2} e^{a}_{m} = \pmatrix{  (R + \epsilon r
cos\vartheta)  & 0 \cr 0 &  r }
\end{eqnarray}
and
\begin{eqnarray}
\label{row_comA3} e^{m}_{a} = \pmatrix{ (R + \epsilon r
cos\vartheta)^{-1}  & 0 \cr
 0 & r^{-1} } \; .
\end{eqnarray}

The Christoffel symbols are equal to
\begin{eqnarray}
\label{row_comA4} \Gamma^{1}_{12} = \frac{- \epsilon r sin
\vartheta} { R + \epsilon r cos \vartheta} \, \, , \, \,
\Gamma^{2}_{11} = \frac{ R + \epsilon r cos \vartheta}{r} \,
\epsilon \, sin \vartheta \, .
\end{eqnarray}
The Ricci tensor $R_{mn}$ components are different from zero
\begin{eqnarray}
\label{row_comA5} R_{11} = \frac{ \epsilon}{ r} cos \vartheta ( R
+ \epsilon r cos \vartheta ) \, \, , \, \, R_{22} = \frac{
\epsilon r cos \vartheta}{R + \epsilon r cos \vartheta}
\end{eqnarray}
which makes the curvature scalar ${\cal R}$ equal to
\begin{eqnarray}
\label{row_comA6} {\cal R} = \frac{ 2 \epsilon cos \vartheta} {r
( R + \epsilon r cos \vartheta )} \, \, .
\end{eqnarray}
Let us notice that only if $\epsilon \rightarrow 0 , {\cal R}
\rightarrow 0 $ we get the flat space. On the other hand, if $R
\rightarrow 0$ and $\epsilon \rightarrow 1$, we obtain ${\cal R}
\rightarrow 2/r^{2}$, similar as for the sphere.

\section*{Acknowledgments}
This work has been supported by the Department of Field Theory
and Particle Physics, Institute of Physics, University of Silesia.

\vspace {40mm}

\section*{Figure captions}

{\bf Figure~1}

The potential $\tilde{V}(\chi)$ for a toy model in scalar field
theory.

\end{document}